\documentclass{optica-article}

\journal{opticajournal} 

\articletype{Research Article}

\usepackage{lineno}
\usepackage{amsmath}
\usepackage{bm}

\begin{document}

\title{Free-induction-decay magnetic field imaging with a microfabricated Cs vapor cell}

\author{Dominic Hunter,\authormark{1,*} Chris Perrella,\authormark{2} Allan McWilliam,\authormark{1} James P. McGilligan,\authormark{1} Marcin Mrozowski,\authormark{1} Stuart J. Ingleby,\authormark{1} Paul F. Griffin,\authormark{1} David Burt,\authormark{3} Andre N. Luiten,\authormark{2} and Erling Riis\authormark{1}}

\address{\authormark{1}Department of Physics, SUPA, University of Strathclyde, Glasgow G4 0NG, United Kingdom\\
\authormark{2}Institute for Photonics and Advanced Sensing (IPAS), and School of Physical Sciences, University of Adelaide, South Australia 5005, Australia\\
\authormark{3}Kelvin Nanotechnology, University of Glasgow, Glasgow G12 8LS, United Kingdom}

\email{\authormark{*}d.hunter@strath.ac.uk} 


\begin{abstract*} 
Magnetic field imaging is a valuable resource for signal source localization and characterization. This work reports an optically pumped magnetometer (OPM) based on the free-induction-decay (FID) protocol, that implements microfabricated cesium (Cs) vapor cell technology to visualize the magnetic field distributions resulting from various magnetic sources placed close to the cell. The slow diffusion of Cs atoms in the presence of a nitrogen (N$_{2}$) buffer gas enables spatially independent measurements to be made within the same vapor cell by translating a $175\,\mu$m diameter probe beam over the sensing area. For example, the OPM was used to record temporal and spatial information to reconstruct magnetic field distributions in one and two dimensions. The optimal magnetometer sensitivity was estimated to be 0.43\,pT/$\sqrt{\mathrm{Hz}}$ within a Nyquist limited bandwidth of $500\,$Hz. Furthermore, the sensor's dynamic range exceeds the Earth's field of approximately $50\,\mu$T, which provides a framework for magnetic field imaging in unshielded environments.

\end{abstract*}

\section{Introduction}
Magnetic imaging encompasses a diverse range of research fields including biomedical science \cite{yang2021new, alem2015fetal, colombo2016four}, current density imaging \cite{hu2020sensitive, bason2022non}, and magnetorelaxometry in magnetic nanoparticles \cite{richter2010magnetorelaxometry, jaufenthaler2021pulsed}, to name a few. Such applications require sensitive instrumentation capable of resolving the spatial and temporal properties of the magnetic sources of interest. For example, living biological specimens have already been imaged at nanometer resolution with diamond nitrogen-vacancy (NV) centers under ambient conditions \cite{balasubramanian2008nanoscale,le2013optical}. The placement of NV centers in close proximity to the sample can greatly benefit such measurements; however, achieving competitive sensitivities is difficult without extensive effort toward fabricating crystals with low impurities \cite{taylor2008high}. Alkali-vapor based radio-frequency magnetometers are commonly implemented for defect detection in metallic objects \cite{deans2016electromagnetic, bevington2019enhanced}, a capability particularly suited for industrial and defense applications. These devices can reach fT-level sensitivities \cite{savukov2005tunable, lee2006subfemtotesla, chalupczak2012room}, especially when operated at higher frequencies where $1/f$ technical noise contributions are suppressed, e.g., laser frequency and intensity drifts. Despite their highly tunable bandwidth, the narrow magnetic resonance linewidth necessary for achieving high sensitivity operation limits the dynamic range. Thus, active or passive field compensation is widely used, inhibiting their compatibility in unshielded environments. \\   
\indent Recent years have seen emphasis placed on miniaturizing OPMs, particularly those operating in the spin-exchange relaxation-free (SERF) regime \cite{osborne2018fully, kitching2018chip}. These devices are already commercially available and have been widely used in the development of magnetoencephalography (MEG) applications; however, they require extensive magnetic shielding in order to function due to their limited dynamic range. Additionally, OPM array density is often constrained by crosstalk between neighbouring sensors, although recent advancements in compact bi-planar coil design and fabrication techniques can aid in alleviating these issues \cite{tayler2022miniature}. Therefore, most OPM-based MEG experiments provide centimeter-scale spatial resolution as each module is self-contained and requires field modulation to operate \cite{boto2018moving, tierney2019optically}. Deploying multiple sensors within a single vapor cell could be advantageous in such applications to increase the spatial resolution \cite{kim2019magnetocardiography}. This method requires fewer vapor cells and allows one to perform gradiometry within the same cell resulting in high common-mode noise rejection \cite{borna2020non}. \\
\indent The OPM modality adopted in this work is based on the well-established FID measurement protocol \cite{hunter2018free, hunter2018waveform, grujic2015sensitive}, as this technique offers distinct benefits toward magnetic imaging applications. For example, as a total-field sensor, its dynamic range extends beyond the Earth's field ($> 50\,\mu$T) whilst maintaining $\mathrm{fT/\sqrt{Hz}}$ sensitivities competitive with zero-field OPM strategies \cite{gerginov2020scalar, limes2020portable, lucivero2022femtotesla}. Furthermore, the sensor bandwidth can extend across a wide range (several kHz), and is highly tunable given the flexibility in the digital signal processing (DSP) techniques that can be used to extract the magnetic field information \cite{hunter2018waveform, wilson2020wide}. This is in contrast to SERF sensors whose bandwidth is typically limited to below $200\,$Hz \cite{osborne2018fully}, owing to the long coherence times needed to reach optimal sensitivity. A major advantage of the FID modality lies in its accuracy \cite{hunter2022accurate}, as the pumping light is switched off during detection which significantly reduces the fictitious magnetic field generated by light shifts. These low systematics reduce the potential sensor background in magnetic imaging experiments. Moreover, the FID approach is extremely robust as it can easily operate in a free-running mode without the use of feedback loops. This enables direct measurement of the Larmor frequency with no prerequisite knowledge of magnetic field required.  \\
\indent Performing spatially independent measurements within a single microfabricated vapor cell for high resolution imaging of magnetic field distributions is a relatively unexplored area; in particular, using the FID modality with a single photodetector in high bias field conditions. Magnetic imaging and source localization is an active field that has seen numerous developments in recent years \cite{taue2010development, taue2020signal}. Spatial resolution is often limited by the, typically cm-scale, sensor footprint \cite{limes2020portable, osborne2018fully}. Higher spatial resolution of $216\,\mu$m has been achieved using a SERF magnetometer employing a digital micromirror device (DMD) \cite{fang2020high}; however, the dynamic range and bandwidth are limited using this OPM mode which places restrictions on unshielded operation. This work explores an alternative approach to magnetic field mapping with a FID-based OPM utilizing micro-electro-mechanical-systems (MEMS) vapor cell technology. The sensor demonstrates competitive $\mathrm{fT/\sqrt{Hz}}$-level sensitivities over a broad dynamic range which provides a platform for high precision magnetic imaging in Earth’s field conditions. This is an extremely desirable characteristic in many applications to reduce, or eliminate, the need for magnetic shielding which has clear value for future commercialization. Furthermore, the sensor's bandwidth capabilities enables resolution of both the spatial and temporal properties of magnetic sources that could extend past the kHz regime \cite{wilson2020wide}. It is also possible to apply this OPM modality to compressed sensing (e.g., using Hadamard patterns \cite{zhang2017hadamard}) with mature technology such as a DMD, providing automation and faster imaging speeds which would facilitate system integration into industrial applications. This is particularly relevant given the customizable cell geometries that can be employed using novel MEMS cell fabrication techniques for improved optical access \cite{dyer2022micro}. Additionally, MEMS cells are more cost-effective and can be manufactured in higher quantities compared to glass-blown cell technology. \\
\indent This study investigates the effect of various current configurations in proximity to the cell by translating the probe beam across the sensing area and recording the magnetic field experienced by the alkali spins. In contrast to paraffin coated vapor cells, in which atoms sample the entire cell volume during measurement, this work exploits the slow alkali spin diffusion inherent to cells with buffer gas. First, the magnetometer's response is verified by applying a well-defined gradient across the sensing volume. Additionally, higher complexity magnetic field distributions produced by current flowing through a copper wire configured in different orientations are compared to theoretical expectations based on the Biot-Savart law. Furthemore, the sensor was able to track the full temporal and spatial characteristics of a magnetic source driven by an oscillating current, constructing a two-dimensional magnetic field image in a bias field of approximately $50\,\mu$T. 

\section{FID Magnetometry}
\subsection{Experimental Setup}
\begin{figure}[b]
	\centering	
        \includegraphics[scale = 1]{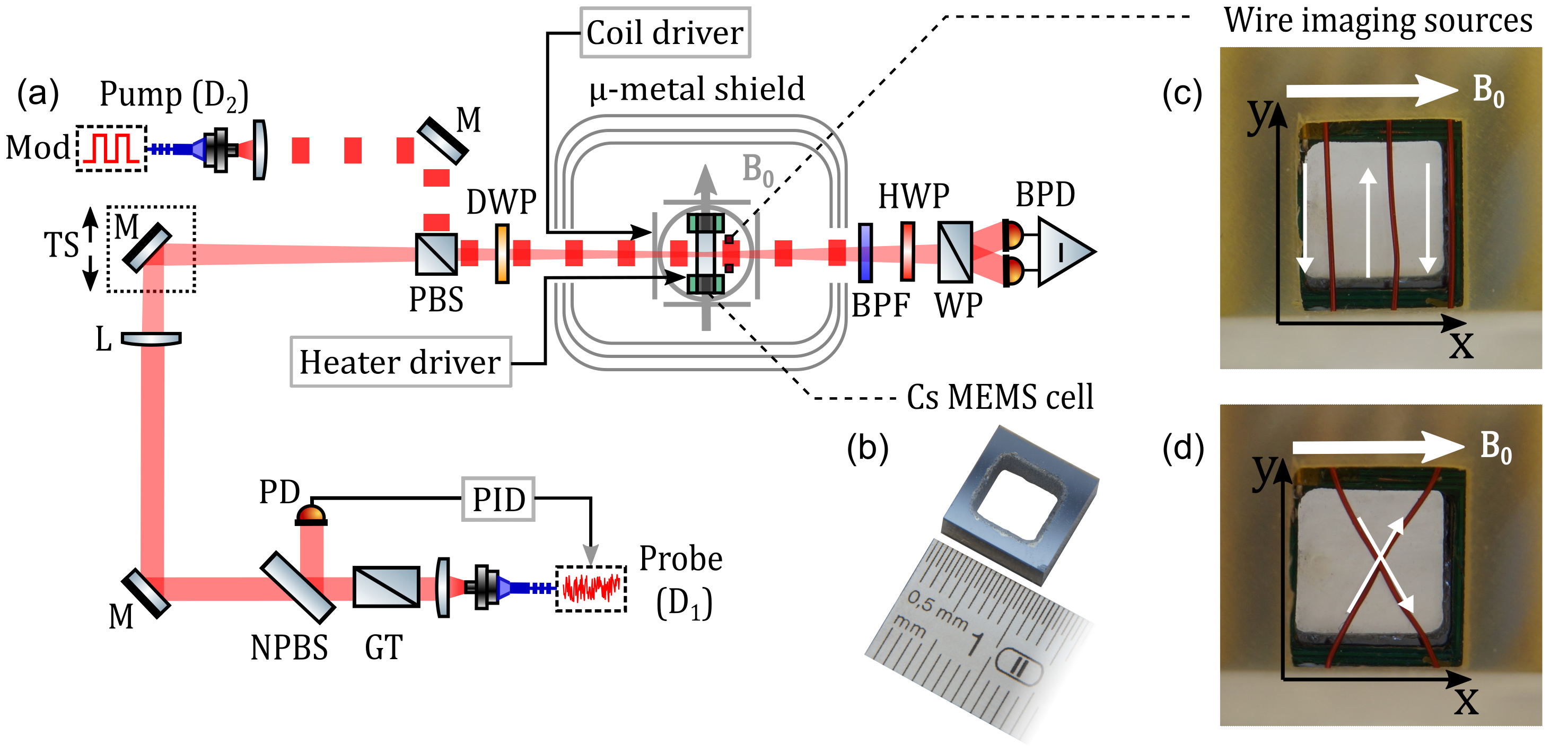}
	\caption{(a) Simplified experimental setup: GT, Glan-Thompson polarizer; NPBS, non-polarizing beamsplitter; PD, photodiode; M, mirror; L, lens; TS, translation stage; PBS, polarizing beamsplitter; DWP, dual-wavelength waveplate; HWP, half-wave plate; BPF, bandpass filter; WP, Wollaston prism; BPD, balanced photodetector. (b) Cs MEMS vapor cell with approximately $220\,$torr N\textsubscript{2} buffer gas and $6 \times 6 \times 3\,$mm$^3$ cavity dimensions. (c, d) Wire configurations used as magnetic sources for one- and two-dimensional (1D and 2D) imaging, respectively. The white arrows indicate the positive current flow paths.}
	\label{Experimental setup}
\end{figure}
A schematic representation of the experimental setup is depicted in Fig.~\ref{Experimental setup}(a). The sensor head consists of a Cs MEMS vapor cell, shown in Fig.~\ref{Experimental setup}(b), formed from a glass-silicon-glass anodically bonded stack. The 3~mm thick silicon wafer is water-jet cut to realize internal dimensions of $6\times6\times3\,\mathrm{mm^3}$ \cite{dyer2022micro}. Following the first glass-silicon bond, an aqueous solution of cesium azide (CsN$_3$) is deposited within the cells. The final bond takes place under vacuum at a pressure of $10^{-5}\,$torr. Thus, the back-filled N$_{2}$ environment is the dominant contributor to the overall buffer gas pressure inside the cell. Following decomposition of the azide under ultra-violet light, the cell measures a total buffer gas pressure of approximately $220\,$torr. This was measured from the collisional broadening and shift in the optical spectrum with respect to a Cs reference cell \cite{hunter2018free}. \\
\indent The Cs MEMS cell is mounted between a set of printed circuit boards (PCBs) used to resistively heat the vapor to a temperature of $88\,^{\circ}$C. This is performed by applying current through the heating element during the optical pumping period, which is rapidly switched to approximately zero at readout. The change in temperature during the power-off period was measured to be lower than the resolution of the temperature monitor ($\approx 60\,$mK), and consequently has no significant impact on the magnetometer signal. The stray field reaches a steady state of $\sim 135\,$fT a few $\mu$s after demagnetization, limited by the leakage current of the MOSFET in the heater driver circuitry. In order to suppress magnetic technical noise from the environment, the sensor head is enclosed inside a three-layer $\mu$-metal shield. It is housed within a set of three-axis field coils that are driven by a highly stable programmable current driver to provide complete control of the magnitude and direction of the bias field \cite{mrozowski2023ultra}. Each magnetic source used for imaging was also powered by the same current source except in instances where modulation was required. In this case, a low-noise waveform generator (Keysight 33600A) was used in series with a resistor with $15\,$ppm temperature stability. \\
\indent Optical pumping and probing were performed using separate co-propagating laser beams tuned to different atomic transitions. The pump laser (LD852-SEV600) is single frequency and can produce up to $600\,$mW of optical power. It is set to be resonant with the $F = 3 \longrightarrow F^{\prime}$ transition on the Cs $\mathrm{D_2}$ line. The optical line is collisionally broadened by the N$_{2}$ buffer gas to a full width at half maximum (FWHM) linewidth of approximately $3.7\,$GHz. This transition is used to optically pump the atomic population from the $F = 3$ hyperfine ground state such that these atoms can subsequently contribute to the signal \cite{schultze2015improving}, while partially suppressing spin exchange through light narrowing. An additional re-pump laser tuned to the $F = 4 \longrightarrow F^{\prime}$ transition would generate more spin polarization at the expense of greater system complexity, given the optical linewidth is smaller than the ground state hyperfine splitting \cite{scholtes2011light}. The pump light was modulated using an acousto-optic modulator (AOM) with approximately $65\,$mW peak optical power available, after beam conditioning and a fiber-coupling stage, which irradiates the vapor cell over a $3.1\,$mm ($1/\mathrm{e}^2$) beam diameter. The pump light intensity is modulated at $1\,$kHz and is maintained at maximum throughout the optical pumping phase. It is subsequently switched off for spin readout with the extinction ratio of the AOM reducing the residual pump light to $<10\,\mu$W when off, limiting interaction with the atoms during detection. \\
\indent The spatial resolution of the magnetometer is governed by the waist of the probe beam which was focused down to a diameter ($1/\mathrm{e}^2$) of $175\,\mathrm{\mu{m}}$, resulting in an elevated optical intensity within the vapor cell. The Rayleigh length ($Z_R \approx 27\,$cm) is an order of magnitude greater than the cell thickness thus intensity gradients along the propagation axis are considered to be negligible. The probe laser (DBR895PN) frequency was $60\,\mathrm{GHz}$ blue-detuned from the $F = 4 \longrightarrow F^{\prime}$ transition of the Cs $\mathrm{D_1}$ line to avoid excessive residual optical pumping from broadening the magnetic resonance during detection. The direction of frequency detuning is not significant in this case given that most of the atoms occupy the $F=4$ ground state after optical pumping and the light is far from resonance. The significant detuning has the additional benefit of reducing light shift systematics to $0.5\,\mathrm{pT/\mu{W}}$, determined experimentally, for improved measurement accuracy. This light shift could be reduced further by employing a Ramsey-style detection protocol \cite{hunter2022accurate}. \\
\indent A probe power of around $1\,$mW was chosen to maximize the signal amplitude while maintaining photon shot-noise level performance. The Glan-Thompson polarizer sets a defined polarization for the probe beam by converting polarization noise into amplitude noise, including that originating from the optical fiber. The light is then split equally between the vapor cell and a monitor photodiode with a non-polarizing beamsplitter. The probe intensity is kept constant with active stabilization by adjusting the RF power supplied to an AOM with an analog proportional-integral-derivative controller (SRS SIM960). Prior to illuminating the vapor cell, a polarizing beamsplitter combines the pump and probe light which then traverse a dual-wavelength multi-order waveplate to convert the $852\,$nm light to circular polarization while the $895\,$nm beam remains linear. This optimizes the optical pumping efficiency whilst maximizing optical rotation of the near-resonant probe beam passing through the vapor. 

\subsection{Sensor Performance}
\begin{figure}
  \centering
  \includegraphics[scale = 0.95]{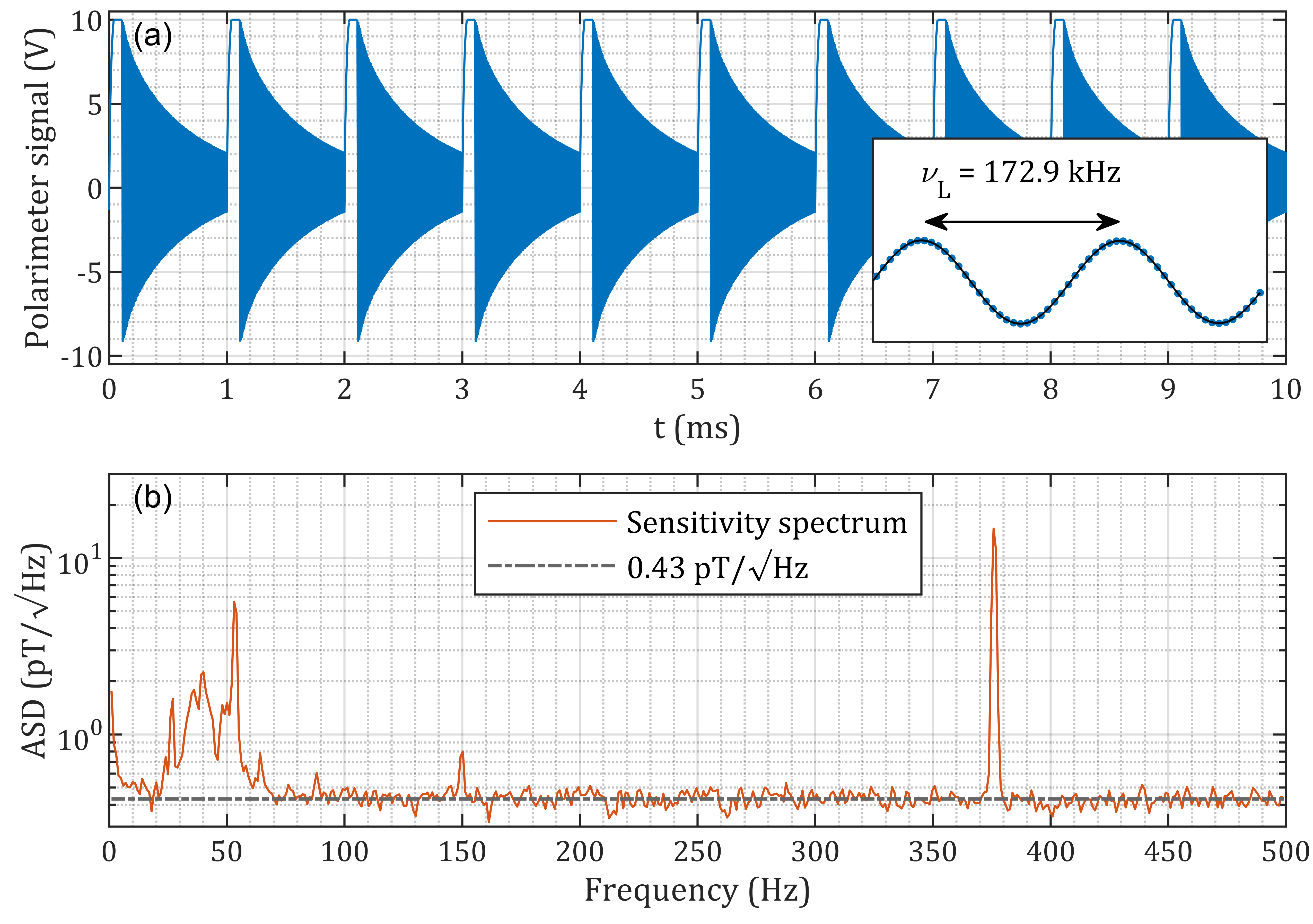}
  \caption{(a) Subsection of a FID signal train captured by the polarimeter in a bias field of $49.4\,\mu$T. The repetition cycle rate was set to $1\,$kHz with the alkali spins optically pumped for around $110\,\mu$s. A Larmor frequency of $172.9\,$kHz was recorded (see inset) by fitting each FID trace to a damped sinusoidal model \cite{hunter2018free}. (b) Optimal magnetic field amplitude spectral density (ASD) computed via Welch's method \cite{welch1967use}, using the extracted Larmor frequency values from a sequence of FID signal trains. The peak observed at $376\,$Hz originates from an oscillating current flowing through a wire used as a magnetic imaging source.}
  \label{FID magnetometer sensitivity}
\end{figure}

Larmor precession induces an oscillating birefringence that is detectable through optical rotation of the linearly polarized probe. A balanced polarimeter detects this rotation and cancels common-mode noise sources such as laser intensity and frequency noise, facilitating photon shot-noise level operation. The bandpass filter spectrally filters residual pump light thus avoiding saturation of the detector. The ensuing analog signal is digitized by a data acquisition system consisting of a Picoscope (model 5444D) sampling at 125 MHz, which is downsampled, by averaging 25 successive data points, to 5 MHz for further processing. Figure \ref{FID magnetometer sensitivity}(a) shows a snapshot of a typical FID signal train captured by the polarimeter in the presence of a $49.4\,\mu$T bias field. The pump-probe cycle repetition period was set to $1\,$ms with close to $110\,\mu$s dedicated to optically pumping the alkali spins. Consequently, the magnetic field data is streamed at a $1\,$kHz sampling rate resulting in a Nyquist limited bandwidth of $500\,$Hz. \\
\indent The Larmor frequency was determined by fitting each FID trace to a damped sinusoidal model with the Levenberg-Marquardt algorithm \cite{hunter2018free}. The inset in Fig. \ref{FID magnetometer sensitivity}(a) shows two oscillations from a fitted FID signal, resulting in a Larmor frequency of $172.9\,$kHz. An optimized magnetometer sensitivity of $0.43\,$pT/$\sqrt{\mathrm{Hz}}$ was calculated from the sensitivity spectrum shown in Fig. \ref{FID magnetometer sensitivity}(b). This spectrum was computed using Welch's method \cite{welch1967use}, by averaging the spectra of magnetic field time series generated from 40 subsequent FID signals trains. In this way, a cleaner magnetic spectrum is gained providing a more precise noise floor estimation. An order of magnitude increase in sensitivity performance is shown compared to previous work \cite{hunter2018free}. This is attributed to the addition of an independent high power pump beam and thicker vapor cells, enabling higher signal amplitudes and longer spin coherence times. The peak observed at $376\,$Hz was produced by sending current through a single wire configured in the manner depicted in Fig. \ref{Experimental setup}(d), later used as a magnetic source for imaging. Technical noise peaks can also be seen between $20-60\,$Hz as a result of magnetic noise in the environment penetrating the $\mu$-metal shield. \\
\indent The magnetometer's spatial resolution is primarily influenced by the probe beam waist within the interrogation region, with the lower bound constrained by the diffraction limit. However, the role of spin diffusion should also be considered which is dictated by the buffer gas pressure inside the cell. One can consider a buffer gas vapor cell as an array of locally independent sensors whose size is equivalent to the distance travelled by the atoms during a measurement. This is known as the diffusion crosstalk-free distance which can be estimated as $\Delta{x} = \sqrt{2DT}$, where $D$ is the diffusion constant calculated for Cs atoms in a N$_{2}$ atmosphere of a specific pressure and temperature, and $T$ is the measurement time \cite{horsley2015high,dong2019spin}. This yields close to $300\,\mu$m when taking into account the $220\,$torr buffer gas pressure of the cell and the readout period of a single FID cycle. This is opposed to paraffin coated cells, in which the atoms sample the entire cell volume during the measurement interval, leading to magnetic gradient broadening of the measured resonance \cite{pustelny2006influence}. By contrast, the coherence time of buffer gas cells is affected by magnetic field variations local to the probe beam within $\Delta{x}$. \\
\indent The probe beam diameter is relatively close to the diffusion length and consequently lies in an intermediate regime in terms of the atomic spin noise profile \cite{lucivero2017correlation}. However, the spin projection noise contribution is negligible as it is an order of magnitude lower than the magnetometer noise floor, which is predominantly limited by photon shot-noise \cite{lucivero2014shot}. The diffusion length can also be reduced by shortening the measurement time, to the detriment of sensitivity, by analysing subsections of the FID signal using, for example, a Hilbert transform \cite{wilson2020wide, hunter2022accurate, ingleby2022digital}. Furthermore, increasing the buffer gas pressure can also lower $\Delta{x}$ by slowing the rate of spin diffusion. \\
\indent The elevated rate of spin destruction collisions at higher buffer gas pressures negatively impacts the spin relaxation rate, and thereby sensor precision based on Cram{\'e}r-Rao lower bound predictions \cite{hunter2018free}. The optimal buffer gas pressure that minimizes the total spin relaxation rate occurs when the rate of cell wall collisions is roughly equal to the buffer gas collision rate. Based on theoretical estimates \cite{scholtes2014intrinsic}, this minimum occurs at $154\,$torr for the cell geometry considered here with the intrinsic longitudinal and transverse relaxation rates calculated to be $\gamma_{10} \approx 267\,$Hz and $\gamma_{20} \approx 1420\,$Hz, respectively.  

\section{Results}
\subsection{Magnetic Field Gradient Calibration}
\begin{figure}[h!]
  \centering
  \includegraphics[scale = 0.95]{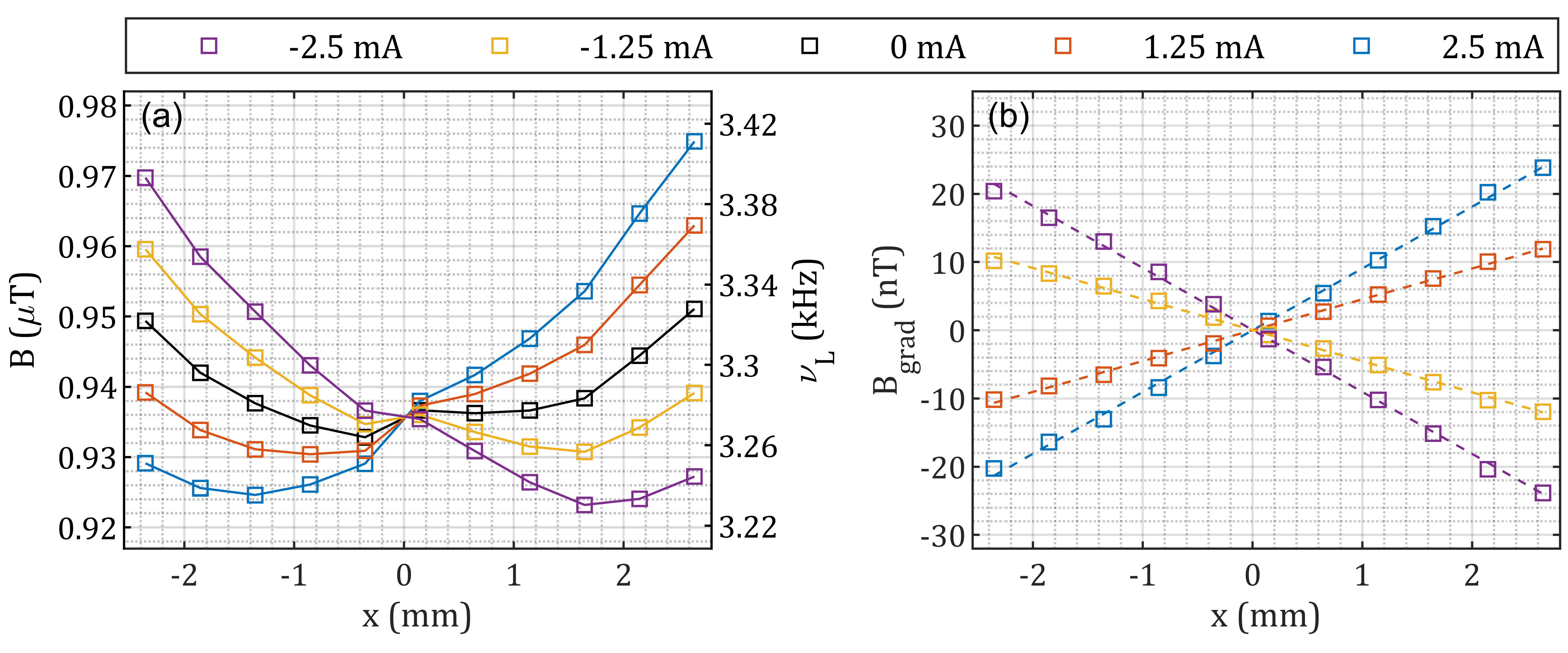}
  \caption{(a) OPM output (squares) when a first-order magnetic field gradient is applied along the $x$-axis using various coil supply currents as noted in the legend. The black data points represent the sensor background with no current supplied to the coil. The solid lines serve as a guide for the eye. (b) Measured magnetic field gradients (squares) at each supply current after background subtraction, and associated linear fits (dashed lines). The ratios between the experimental and theoretically predicted (see Eq. \ref{eqn: gradient}) gradients, in ascending order of current, were 0.934, 0.931, 0.943, and 0.937. All markers are larger than the associated error bars.}
  \label{Gradient magnetic field distribution}
\end{figure}

An initial evaluation of the sensor head's suitability for magnetic imaging applications was conducted by measuring a well-defined first-order magnetic field gradient along the $x$-axis. Additionally, an offset bias field, $B_0 \approx 0.95\,\mu\mathrm{T}$, was applied along the same axis. The gradient field was produced by a single-turn ($N=1$) counter-wound coil pair separated by $s \approx 19.7\,$mm, each with radius $R \approx 16\,$mm. To demonstrate the effect of different gradient fields on the sensor output, the current supplied to the coil was varied between $\pm\,2.5\,$mA. The probe light was reflected by a mirror mounted on a translation stage as seen in Fig. \ref{Experimental setup}(a), enabling adjustment of the beam position along the $x$-axis of the vapor cell with $0.25\,$mm resolution. The average magnetic field obtained from the measured Larmor frequency values of a $1\,$s FID signal train was computed at several $x$-axis positions as shown in Fig. \ref{Gradient magnetic field distribution}(a). 
\\
\indent The observations are compared to theoretical predictions based on calculating the field strength, $B(x)$, at a position, $x$, from the midpoint between two identical parallel plane coils whose current flows are in opposite directions. The gradient field distribution can be determined from the following analytical expression,
\begin{equation}
\label{eqn: gradient}
    \begin{split}
        B(x) = \frac{\mu_0 N I R^{2}}{2} & \biggl(\left[R^2 + (x-s/2)^2\right]^{-3/2} -  \left[R^2 + (x+s/2)^2\right]^{-3/2}\biggr),  
    \end{split}
\end{equation}
\noindent where $\mu_0$ is the magnetic permeability, $N$ is the number of coil turns, $I$ is the current, $R$ is the coil radius, and $s$ is the coil separation. \\
\indent Most of the field measured by the magnetometer is produced by $B_{0}$. There are additional minor background variations observed in the absence of current, as denoted by the black data points in Fig. \ref{Gradient magnetic field distribution}(a). These can arise from multiple sources, including gradients produced when applying the bias field, and optical pumping effects. For example, optical pumping of the alkali spins within the cell is inhomogeneous due to the finite waist and Gaussian profile of the pump beam over the imaging area. Signals are still observed even at the furthest edges of the cell due to the high optical power of the pump beam; however, there is a reduction in signal amplitude, and consequently sensitivity, when the probe beam samples these lower intensity regions. This is a signature that the atomic population distribution over both hyperfine ground states is a function of position within the vapor cell due to variations in optical pumping efficiency. As a result, the Larmor frequency measurements will be weighted by these populations since the ground states have slightly different gyromagnetic factors. This is a source of heading error which can be compensated for analytically in the high spin polarization limit \cite{lee2021heading}.   \\
\indent Figure \ref{Gradient magnetic field distribution}(b) shows the measured magnetic field gradients after subtracting the sensor background. As expected, a linear relationship is clearly evident between the observed magnetic field and $x$-axis position of the probe beam for each supply current. The experimental gradients coincide with the theoretical estimates to within $7\,\%$, verifying the OPMs ability to spatially resolve magnetic field distributions. The observed offset likely results from inaccuracies in the theoretical predictions, since they are extremely sensitive to minor deviations in coil geometry. This is validated by considering the ratio between the experimental and theoretical gradients for each current case which are in agreement with each other to within $\pm\,0.6\,\%$.  

\subsection{1D Magnetic Field Imaging}
\begin{figure}
  \centering
  \includegraphics[scale=0.95]{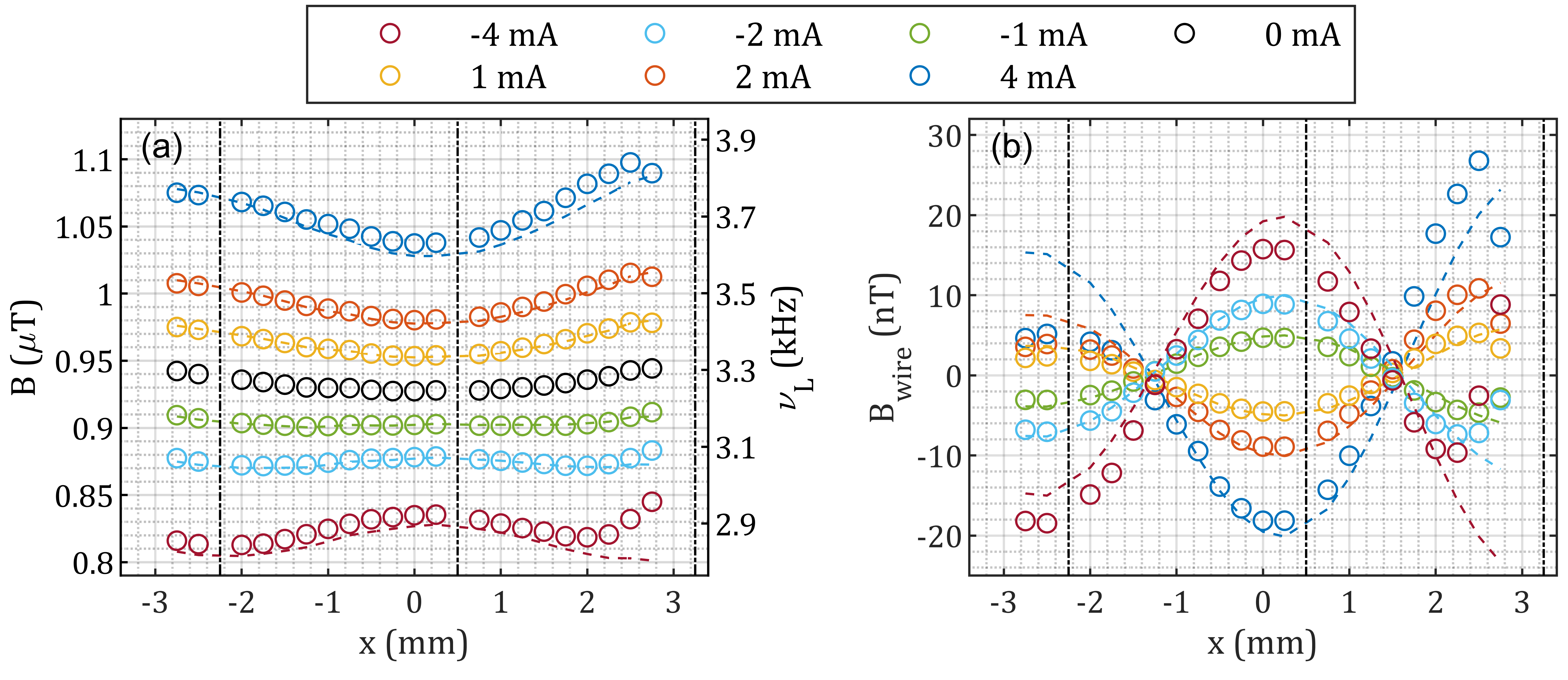}
  \caption{(a) 1D magnetic field mapping of the current configuration depicted in Fig. \ref{Experimental setup}(c). The positions of the copper wire are represented by vertical dashed-dotted lines, with the supply currents noted in the legend. The OPM readings (circles) are compared with theoretical predictions (dashed lines) based on the Biot-Savart law shown in Eq. \ref{eqn:BS}. (b) Recorded magnetic field distribution and associated theoretical model after subtraction of the the sensor background and subsequent mean values. The slight asymmetry likely stems from deformations in the wire and interference produced by the external feedthroughs. All markers are larger than the associated error bars.}
  \label{Wire 1D scan}
\end{figure}
A local magnetic field source was placed close to the vapor cell to generate a more complex field distribution along the $x$-axis. A copper wire (32 awg) was positioned approximately $4.5\,$mm from the center of the vapor cell in the configuration depicted in Fig. \ref{Experimental setup}(c), consisting of three adjacent wire segments spaced approximately $2.75\,$mm apart with alternating current flows. It was placed on the light exit face of the cell so that the pump and probe beams could interact uninterrupted with the vapor. The locations of the wires are indicated by the vertical dashed-dotted lines in Fig. \ref{Wire 1D scan}, showing the recorded data. No data could be collected at these positions as the probe beam was obstructed by the wire and did not reach the balanced photodetector. \\
\indent The magnetometer readings are denoted by circles in Fig. \ref{Wire 1D scan}(a) for various currents flowing through the wire ranging between $\pm\,4\,$mA. The sensor background measured at $0\,$mA (black circles) matched closely to the observations in Fig. \ref{Gradient magnetic field distribution}(a). The dominant contribution is the magnetic field gradient generated as a byproduct of the bias field. It can be seen that positive current flow in the central wire segment generates a magnetic field that opposes $B_{0}$, whereas the wires near the edges produce fields that add constructively. This is to be expected given the geometries indicated in Fig. \ref{Experimental setup}(c). In addition, there is an overall increase in the measured field with respect to the sensor background as the wire segments at the edges dominate. Also, as anticipated, reversing the direction of the current inverts the observed magnetic field distribution and thus lowers the overall field. \\
\indent This behaviour is consistent with theoretical predictions (dashed lines) derived from the Biot-Savart law,
\begin{equation}\label{eqn:BS}
    \bm{B}(\bm{r}) = \frac{\mu_{0}}{4 \pi} \int \frac{I\,d{\bm{l}} \times \hat{\bm{r}} }{\bm{r} \cdot \bm{r}},
\end{equation}
where $I$ is the current along increment $d{\bm{l}}$, and $\bm{r}$ is the position with unit vector $\hat{\bm{r}}$. The contributions are approximated as three infinite straight wires $\bm{B}(\bm{r})_{1}$, $\bm{B}(\bm{r})_{2}$, and $\bm{B}(\bm{r})_{3}$, calculated using Eq. \ref{eqn:BS} for arbitrary positions within the cell and averaged over the length of the vapor cell that the probe beam passes through, yielding $\overline{B}$,
\begin{equation}\label{eqn:BAvg}
    \overline{B}(x,y) = \frac{1}{L}\int_{0}^{L} \left\|\bm{B}(\bm{r})_1+\bm{B}(\bm{r})_2+\bm{B}(\bm{r})_3\right\|\,d\,z.
\end{equation}
The average is made as the probe beam samples the atomic vapor over the length, $L = 3\,$mm, of the cell with the atoms experiencing different field strengths along this path which our observation averages over. This effect of averaging could be minimized using vapor cells with shorter optical path lengths such as that described in Ref. \cite{hunter2018free}. The wire segments were assumed to be $4.5\,$mm from the center of the cell along the $z$-axis which aligns with experimental conditions. \\
\indent It can be seen from Fig. \ref{Wire 1D scan} that there is a slight deviation of the experimental data from the theoretical trends, particularly toward the right edge of the cell, which is likely caused by interference from the external connecting wire. This is particularly evident from the asymmetry observed in Fig. \ref{Wire 1D scan}(b) where the sensor background and subsequent mean values have been subtracted, and more succinctly highlights the summation of fields produced by adjacent wire segments. It should also be noted that the theoretical predictions do not account for deformations, the most noticeable of which is in the central wire segment as seen in Fig. \ref{Experimental setup}(c). 

\subsection{2D Magnetic Field Imaging}
To facilitate 2D magnetic imaging, the mirror was tilted vertically to translate the beam along the $y$-axis in increments of $1\,$mm. The angle of the beam traversing the cell is insignificant given the $3\,$mm optical path compared to the distance ($\sim 0.5\,$m) between the vapor cell and the scanning mirror. The probe beam's position along the $y$-axis was determined using a CMOS camera temporarily placed in front of the cell. The probe beam's position within the cell was estimated by measuring the respective distances between the camera and the vapor cell from the scanning mirror. Careful alignment and measurement of beam position relative to the vapor cell ensures measurement accuracy and repeatability. This aspect could be improved further through automated translation (e.g., using a DMD). The magnetic source is depicted in Fig. \ref{Experimental setup}(d), which invokes magnetic field structure over two dimensions. The wire was placed roughly $4.5\,$mm from the center of the $3\,$mm thick vapor cell, consistent with the 1D scan. \\
\begin{figure}[ht]
	\centering
	\includegraphics[scale=0.95]{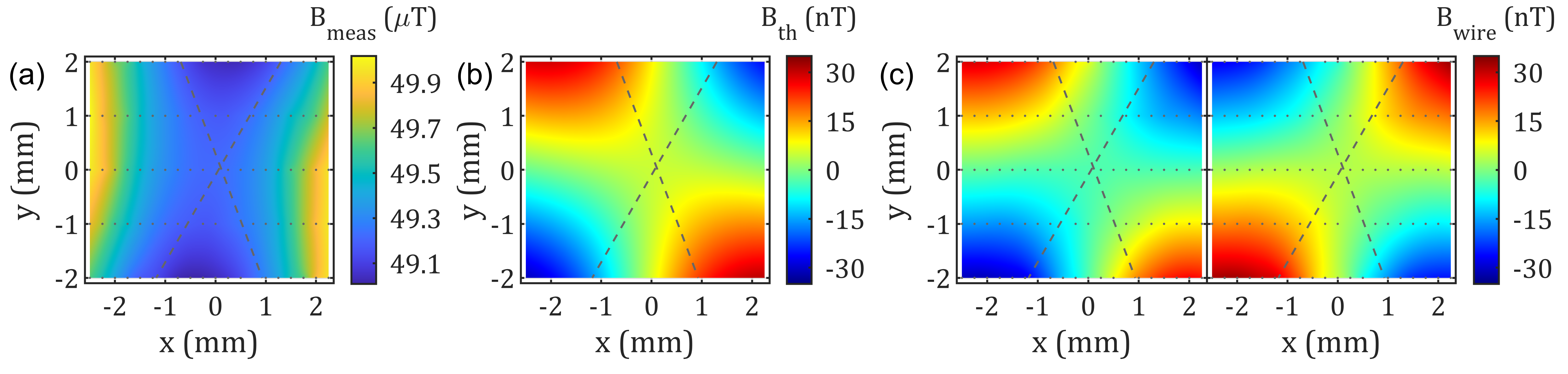}
	\caption{(a) Magnetometer output at each probe beam location (grey dots) with $3.08\,$mA current flowing through the wire. The majority of the magnetic field variation is a byproduct of $B_{0}$ applied along the $x$-axis. (b) Theoretical magnetic field distribution based on the Biot-Savart law (see Eq. \ref{eqn:BS}) for the wire configuration depicted in Fig. \ref{Experimental setup}(d) at a current of $3.08\,$mA. (c) Distributions produced at currents of $3.08\,$mA (left) and $-3.13\,$mA (right) after background subtraction. Dashed lines indicate the approximated wire positions. Natural-neighbour interpolation was used throughout to enhance data visualization \cite{sibson1981brief}.}
	\label{Cross-wire 2D scan}
\end{figure}
\indent The magnetometer output over the imaging area is shown in Fig. \ref{Cross-wire 2D scan}(a) in the presence of a bias field, $B_{0} \approx 49.5\,\mu$T, along the $x$-axis and $\mathrm{3.08\,}$mA of current passed through the wire. This increase in bias field highlights the extensive dynamic range exhibited by the device in the context of magnetic imaging. It can be seen that a significant magnetic field gradient is produced as a byproduct of the bias field. Figure \ref{Cross-wire 2D scan}(c) portrays the magnetic field distributions, after background subtraction, produced by passing static currents of $-3.13\,$mA (right) and $3.08\,$mA (left) separately through the wire. It can be seen by comparing these distributions that reversing the current flow inverts the magnetic field image in the anticipated manner. Adding the distributions yielded a standard deviation of $1\,$nT, which is $1.7\,\%$ of the wire's total field variation over the imaging area. The wire produces a field range contributing to $<0.1\,\%$ of $B_{0}$, and accounts for roughly $6\,\%$ of the total magnetic gradient across the cell observed in Fig. \ref{Cross-wire 2D scan}(a). The behaviour shown in the left distribution in Fig. \ref{Cross-wire 2D scan}(c) is well represented by theoretical expectations based on the Biot-Savart law, depicted in Fig. \ref{Cross-wire 2D scan}(b) for this current configuration. It can be seen that the experimental and theoretical trends both provide similar patterns and extend over a range of approximately $60\,$nT within the imaging area. \\
\begin{figure}[t]
	\centering
	\includegraphics[scale = 0.95]{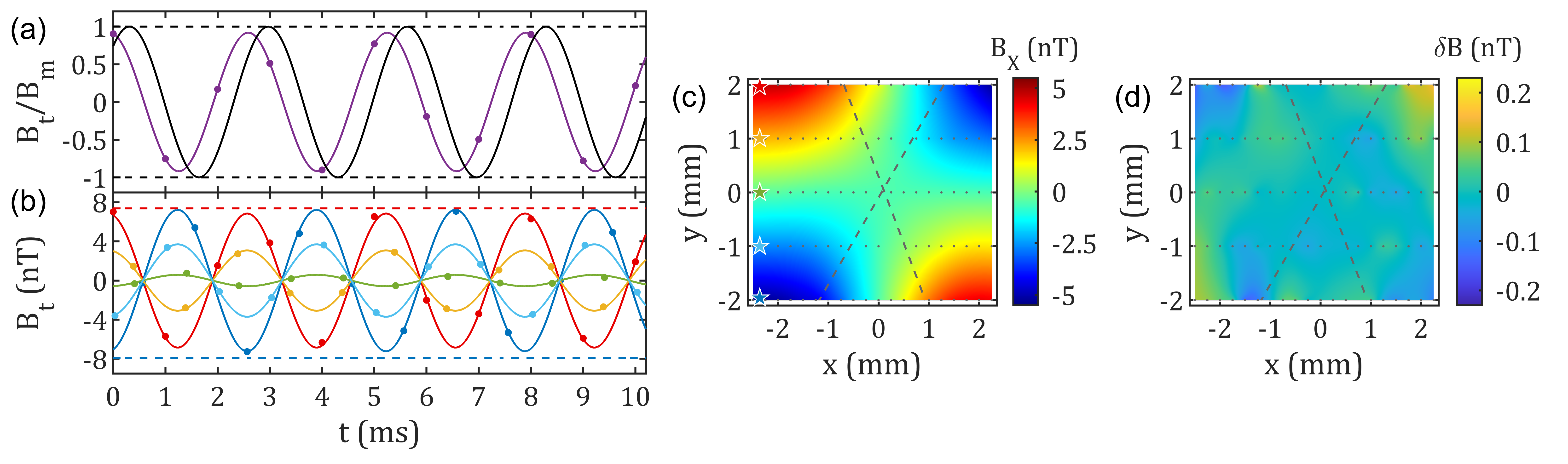}
	\caption{(a) Simulated FID OPM response (purple dots) to a field modulation (black line) of amplitude, $B_{m}$, and corresponding sinusoidal fit (purple line). (b) First $10\,$ms of OPM recordings at different probe beam positions as indicated in (c). Data points are colored according to the stars represented at each location. Horizontal lines illustrate the DC response measured for the top and bottom probe beam positions. (c) Magnetic field distribution produced by a $0.55\,$mA root-mean-square current modulation at $376\,$Hz. Each $1\,$s OPM recording is software demodulated providing the response $B_{X}$. (d) Difference between the DC and oscillating magnetic field components after scaling according to the respective current amplitudes and expected frequency response at $376\,$Hz.}
	\label{Wire 2D scan AC}
\end{figure}
\indent To demonstrate the magnetometer's ability to detect oscillating magnetic sources, a $0.55\,$mA root-mean-square current modulation was superimposed onto the static field mentioned previously at a frequency of $376\,$Hz. The frequency and phase responses of the FID modality are not completely flat within the Nyquist limit and will depend on the applied signal processing strategy. Figure \ref{Wire 2D scan AC}(a) depicts the anticipated magnetometer response to a $376$\,Hz modulation. Simulated FID data was generated using a model that accounts for modulation of the Larmor frequency \cite{hunter2018waveform}. Each simulated FID trace samples the modulation field at a different phase. The frequency, i.e., magnetic field, is subsequently determined using the same fitting technique applied to the experimental traces. For each magnetic field value, the time stamp is set at the beginning of the FID cycle, e.g., the first value starts at $0\,\mathrm{ms}$. The fitting algorithm will define the magnetometer phase and frequency response as the magnetic field changes significantly within a single FID cycle whereas the model assumes a single frequency component. Consequently, the frequency determination depends on which segment of the modulation field is sampled during the readout period and is weighted according to the exponential envelope caused by decoherence. \\
\indent  Figure \ref{Wire 2D scan AC}(b) displays the first $10\,$ms of the $1\,$s magnetic field time series collected at selected probe beam positions, denoted by the stars in Fig. \ref{Wire 2D scan AC}(c). It can be seen that the red data set is representative of the predicted phase and amplitude response. The red and blue dashed lines indicate the DC response at the corresponding beam locations after scaling with respect to the current amplitudes. Probe beam translation has a noticeable effect on the observed signal amplitude which experiences a $\pi$ phase shift as a consequence of the current configuration. Magnetic field data collected at each position was software demodulated to extract the magnetic field amplitude, $B_{X}$, providing the distribution in Fig. \ref{Wire 2D scan AC}(c). It can be seen that the observed magnetic image is consistent with that shown in Fig. \ref{Cross-wire 2D scan}(c), taking into account the current amplitude which is approximately $5.6\,\times$ smaller than the static offset. Figure \ref{Wire 2D scan AC}(d) shows the difference between the DC and oscillating distributions after scaling based on the relative current amplitudes and accounting for the slight magnetometer roll-off. Approximately $0.6\,\%$ fluctuation of the total field variation produced by the wire is observed. \\
\indent Performing these magnetic imaging experiments at higher frequency is advantageous as no background subtraction is required, and there is flexibility in selecting a frequency that resides in a clean part of the noise spectrum. Additionally, it extends the magnetometer's utility to imaging of high frequency magnetic sources. 

\section{Conclusion and outlook}
In summary, the magnetic field imaging capability of a FID magnetometer based on a Cs MEMS vapor cell with approximately $220\,$torr N$_{2}$ buffer gas was demonstrated. This was achieved by translating the readout beam across different regions of the sensor head and measuring the magnetic field at each location. The OPM response was validated using theoretical predictions of the field distributions produced by a gradient field and a wire magnetic source. Two-dimensional magnetic imaging was conducted at a high bias field of approximately $50\,\mu$T. This illustrates the magnetometer's extensive dynamic range and offers a pathway toward magnetic imaging in unshielded environments. Furthermore, the wide bandwidth of the sensor was exploited to simultaneously resolve the temporal and spatial magnetic field components produced by the wire. \\  
 \indent Aligning the probe beam through the cell in a double-pass arrangement would improve this magnetic imaging strategy since it eliminates light obstruction caused by the magnetic source, and elevates the sensitivity through increased optical rotation. This could be achieved by placing a reflector on the back surface of the cell prior to the imaging source, allowing enhanced accessibility of the sensor head toward the magnetic source of interest. Such a sensor would be an invaluable resource in PCB inspection and quality assurance of integrated circuits (ICs) \cite{holzl2012quality, hofer2012analyzing}. The current distribution of an IC has already been measured using NV centers \cite{kehayias2022measurement}, although these devices demonstrate limited sensitivity. The output could be compared to known distributions obtained from functional devices, to rapidly test ICs in a production line setting. The higher precision achievable with atomic magnetometers would be particularly valuable in battery diagnostics through non-invasive current mapping \cite{bason2022non}. These measurements could benefit from more sensitive instrumentation, and have already been performed using zero-field OPMs \cite{hu2020rapid}. These devices demonstrated a $5\,$nT dynamic range, hence require shielding and could only tolerate limited nickel concentrations; a problem that can be overcome using total-field sensors such as that considered in this work.  \\
 \indent There is also potential for improving homogeneity in the optical pumping dynamics across the cell by providing a more uniform pump beam intensity profile. Consequently, this ensures a consistent spatial sensitivity dependence, and reduces the overall systematics that contribute to the sensor background. A simple solution would be to expand the pump beam; however, this would result in a lower peak intensity and excess beam clipping at the edges of the cell, negatively impacting sensor sensitivity. Alternatively, a flat-top beam profile tailored to the cell's dimensions could be implemented, although this requires complex beam shaping and will likely result in optical losses. The most effective solution would be to translate both the pump and probe beams simultaneously by launching them through the same optical fiber for optimal optical pumping efficiency at every position. This would also provide easier system integration as the sensor head and high-power modulated light source can be separated using a modular design. Future experiment iterations will focus on combining this approach with a double-pass geometry to improve both the consistency and optimal sensitivity across the entire image area. \\
 \indent A fixed probe beam is also suitable for magnetic imaging using differential measurements between sensitive regions in a sensor array. For example, this has been accomplished by employing quadrant photodiodes detecting a single fixed probe beam \cite{lucivero2022femtotesla}. In real-world applications, such an approach could be advantageous for background subtraction without prior knowledge of the ambient magnetic field conditions. Image reconstruction speeds could be accelerated using a spatial light modulator (SLM) such as a DMD \cite{fang2020high}, as opposed to the slower process of probe beam translation. In this case, magnetic field distributions would be reconstructed with a single pixel detector, such as a photodetector, by applying spatially varying illumination patterns \cite{zhang2017hadamard}. The photodetector signal collected from each spatial light mode would result in FID data that can be subsequently analysed to extract Larmor frequency information. The magnetic image would be reconstructed by correlating symmetries in the spatial light modes with the corresponding magnetic field data. Accordingly, magnetic images could be formed more efficiently as DMDs can deliver refresh rates in the kHz regime. In this case, the measurement bandwidth would be limited by the repetition rate of the OPM. However, one could circumvent this issue by utilizing alternative DSP techniques, such as a Hilbert Transform \cite{wilson2020wide, hunter2022accurate, ingleby2022digital}, for quicker Larmor frequency extraction. Also, the spatial resolution becomes dependent on how many spatial modes are used. Thus, the probe beam width can be extended without sacrificing spatial resolution in order to enhance sensitivity performance.

\section*{Funding: \normalfont{\footnotesize{Engineering and Physical Sciences Research Council (EP/W026929/1); Innovate UK (ISCF-42186).}}} 

\section*{Acknowledgements: \normalfont{\footnotesize{JPM gratefully acknowledges funding from a RAEng research fellowship. AM was supported by a Ph.D. studentship from the Defence Science and Technology Laboratory (Dstl).}}}

\section*{Disclosures: \normalfont{\footnotesize{The authors declare that there are no conflicts of interest related to this article.}}}

\section*{Data availability: \normalfont{\footnotesize{Data underlying the results presented in this manuscript are available in Ref. \cite{hunter2023Free}.}}}

\bibliography{references}

\end{document}